\documentclass[11pt,dvips]{article}
\usepackage{epsfig,times}
\usepackage[russian,english]{babel}
\usepackage[cp1251]{inputenc}

\begin{document}
\begin{center}
{\bf \Large 
Long-range and short-range multiparticle correlation in collisions of $_{79}Au^{197}$ 10.7 A GeV with photoemulsion nuclei
} 
\vspace{2mm}

{\it \large A.Sh. Gaitinov, I.A. Lebedev, A.I. Lebedeva}

\vspace{2mm}

{\large Institute of Physics and Technology, Almaty, Kazakhstan}

\vspace{2mm}

\end{center}

{\small 
A search and research of long-range and short-range multiparticle correlations in pseudorapidity distribution of secondary particles on the base of analysis of individual interaction of nuclei of  $^{197}Au$ at energy 10.7 AGeV with photoemulsion nuclei, is carried out.
Events with long-range multiparticle correlations (LC), short-range multiparticle correlations (SC) and mixed type (MT) in pseudorapidity distribution of secondary particles, are selected by Hurst method in accordance of Hurst curve behavior. 
These types have significantly different characteristics. At first, they have different parameters of fragmentation. Events of LC type are processes of full destruction of projectile nucleus, in which multicharge fragments are absent. In events of mixed type several multicharge fragments of projectile nucleus are discovered. Secondly, these two types have significantly different multiplicity distribution. Mean multiplicity of LC type events is significantly more than in events of mixed type. 
      On the basis of research of dependence of multiplicity versus target-nuclei fragments number for events of various types it is revealed, that the most considerable multiparticle correlations are observed in interactions of mixed type, which correspond to the central collisions of gold nuclei and nuclei of CNO-group, i.e.  nuclei with strongly asymmetric  volume, nuclear mass, charge, etc. Such events are characterised by full destruction of target-nucleus and disintegration of projectile-nucleus on several multi-charged  fragments.

}

\large

\section{Introduction}

    Search of phase transition of matter from hadron conditions to quark-gluon plasma is one of the important problems of high energy 
physics \cite{1}-\cite{3}. Experimental and theoretical researches in this area traditionally concentrate on studying of interactions of 
nuclei at high energies, creating the best conditions for studying of such problems: high pressures and temperature in reaction volume.

Fluctuations and correlations are signs of phase transition. 
In particular phase transition from quark-gluon phase to hadron phase can lead to essential fluctuations in distributions of 
secondary particles \cite{10}.

The event-by-event analysis of high energy nuclear collisions aims at searching for dynamical fluctuations associated with the 
phase transition of normal nuclear matter to the Quark-Gluon Plasma. 
Fluctuations of thermodynamic quantities provide an unique framework for studying the nature of QGP phase transition and
provide direct insight into the properties of the system created in high energy heavy ion
collisions \cite{10-1},\cite{10-2}.

In the present paper we analyzed experimental pseudorapidity distributions of secondary particles obtained in 10.7 A GeV $ ^{197} Au$ 
interactions with emulsion nuclei \cite{11}.

    Nuclear photoemulsion in comparison with other approaches, which are used for investigation of nuclei interactions, is one of 
the most informative.
At first, it has high spatial resolution. Secondly the nuclear photoemulsion allows to observe an interaction in $4\pi$ geometry 
of experiment. The most of other methods have essential dead zones in which secondary particles are not registered.   

    Besides, the method of nuclear emulsion allows to define charges of fragments and gives the chance to registration rather small 
excited nucleus-targets, and also has no energy threshold of registration of fragments of a nucleus-projectile.

    Moreover, the emulsion technique allows to investigate parameters of fragmentation and multiparticle processes in the same interaction.

Peculiarities of secondary particles distribution we analyzed by Hurst method. 
In work \cite{12} it was shown that the method allows not only to discriminate dynamic multiparticle correlations from statistical ones, 
but also to determine a "force" and "length" of these correlations.

\section{Analysis procedure}

According to the existing notions,
secondary particles, which are "emited" from "interaction volume", $n_s$ have
pseudorapidities, corresponding to a central region of pseudorapidity
distribution. At borders of the distribution, fragments (of the target-nucleus
and projectile-nucleus) bring in considerable contribution. And so, in order
to investigate of pseudorapidity
correlations in distribution of particles from "interaction volume"
we have chosen central pseudorapidity interval $\Delta \eta =  4$.
This interval has been subdivided into $k$ parts with $\delta\eta=\Delta\eta /k$.
By counting the number of particles in each subinterval we arrive at a
sequence $n_i^e$.

A pseudorapidity fluctuation, or the normalized relative deviation
of an individual event from average pseudorapidity distribution
\footnote{It is possible to use also absolute deviation
$\xi_i =  n_i^e/n^e - n_i/n.$} is given by
\begin{equation}
\xi_i = \frac{n_i^e/n^e - n_i/n}{n_i/n}= \frac{n_i^e}{n^e} \;
\frac{n}{n_i} - 1
\end{equation}
where $ n_i^e$ is the number of particles in the i-th bin of an event
with
particles number $n^e$, and $n_i= \sum_en_i^e$ is the total number of
particles
for all events in the i-th bin, and $n= \sum_en^e$ is the total number
of
particles for all events.

As described in our previous work \cite{12} for an investigation of
pseudorapidity correlations we analysed the normalized range
$H(k')= R(k')/S(k')$ (where $R(k')$ and $S(k')$ are a "range" and a standard
deviation, which is calculated by Eqs.(\ref{s})-(\ref{1}), see below)
versus the size of the pseudorapidity interval $d\eta =  k'\delta\eta$,
($1\le k' < k$) using a function
\begin{equation}
H(k')= (a k' )^h
\label{6}
\end{equation}
where $a$ and $h$ are two free parameters and $h$ is the correlation
index (or Hurst index). If the signal $\xi_i$ represents white noise
(a completely uncorrelated signal), then $h= 0.5$.
If $h>0.5$, long-range correlations are in a system \cite{13},\cite{14}.

In our calculations we have used $k= 1024$. 
So, for the sequence $\xi_i$, $1\le i\le k$,
quantities of $R(k)$ and $S(k)$ were calculated by following formulas:
\begin{equation}
S(k)= \left[ \frac{1}{k} \sum^{k}_{i= 1} [\xi_i-<\xi >]^2\right]^{1/2}
\label{s}
\end{equation}
\begin{equation}
R(k)= \underbrace{max X(m,k)}_{1\le m\le k} -\underbrace{min
X(m,k)}_{1\le m\le k}
\end{equation}
where the quantity $X(m,k)$ characterizes the accumulated deviation from
the average
\begin{equation}
< \xi > = \frac{1}{k} \sum^{k}_{i= 1}\xi_i
\end{equation}
for a certain interval $m\delta\eta$,
\begin{equation}
X(m,k)= \sum^{m}_{i= 1} [\xi_i-<\xi >], \quad 1\le i \le m \le k
\label{1}
\end{equation}

Then the sequence $\xi_i $ has been subdivided into two parts: $\xi_i'$,
$1\le
i\le k'= k/2$ and $\xi_i''$, $k'+1\le i\le k$, and the value of
$H(k/2)= R(k/2)/S(k/2)$ was found for each of the two independent
series.
Similarly $\xi_i'$ and $\xi_i''$ have been subdivided further, followed
by the
calculation of $H(k/4)= R(k/4)/S(k/4)$. This subdivision and analysis
procedure
for newly obtained series of $\xi_i$-values is continued until $d\eta >
0.1$.
$H$ corresponding to the same value of $k'$ have been averaged and drawn
on a
log-log scale as a function of $k'$.

\section {Results and discussion}

   On the basis of the described procedure we have calculated for individual events the Hurst index, which was defined by the equation (2).

   In Figure 1 the distribution of events with various values of the Hurst index versus multiplicity of $n_s$ particles, is presented.

\begin{figure}[tbh]
\begin{center}
\includegraphics*[width=0.7\textwidth,angle=0,clip]{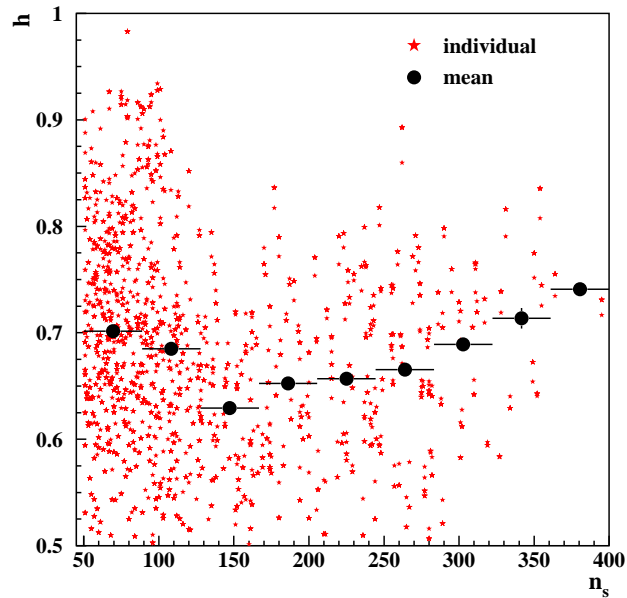}
\caption{\label{fig1} 
The distribution of events with various values of the Hurst index versus  
$n_s$ multiplicity for individual events of interaction of nuclei of  $^{197}Au$ at 10.7 AGeV with photoemulsion nuclei }
\end{center}
\end{figure}

    The analysis of the distribution, which is presented in the Figure 1, finds out essential correlation dependence of the Hurst index versus multiplicity of $n_s$ particles. At first, $h$ index growth at increasing of  $n_s$ in an interval from 150 to 400, is observed. Secondly, at multiplicity from 50 to 150 considerable quantity of events with the high indicator $h> 0.8$, is found out. 
Further all events have been conditionally divided into two classes: usual events (with stochastic pseudorapidity fluctuations) and correlated events (with essential multiparticle correlation). 
The selection of events was made on the basis of Hurst index $h = 0.64$.

       In work \cite{15} it was shown that the criterion $h = 0.64$ corresponds to process in which all secondary particles were produced from two-particle decays. And so, this criterion conditionally divides all experimental set into processes, in which certain dynamic multiparticle correlations are observed, and on events, in which multiparticle correlations are absent. 

   Therefore, if Hurst index was more than $0.64$, then event was referred into group of correlated events. If $h$ was less than 0.64, event has been named uncorrelated (poorly correlated).

For research of possible distinctions in the mechanism of formation of a final pseudorapidity distribution for two types of events, 
we have analysed behaviour of fragments of target-nuclei and projectile-nuclei. 

The most essential differences for two types of interactions are found out for $N_h$ distributions, which are presented in Figure 2.

\begin{figure}[tbh]
\begin{center}
\includegraphics*[width=0.48\textwidth,angle=0,clip]{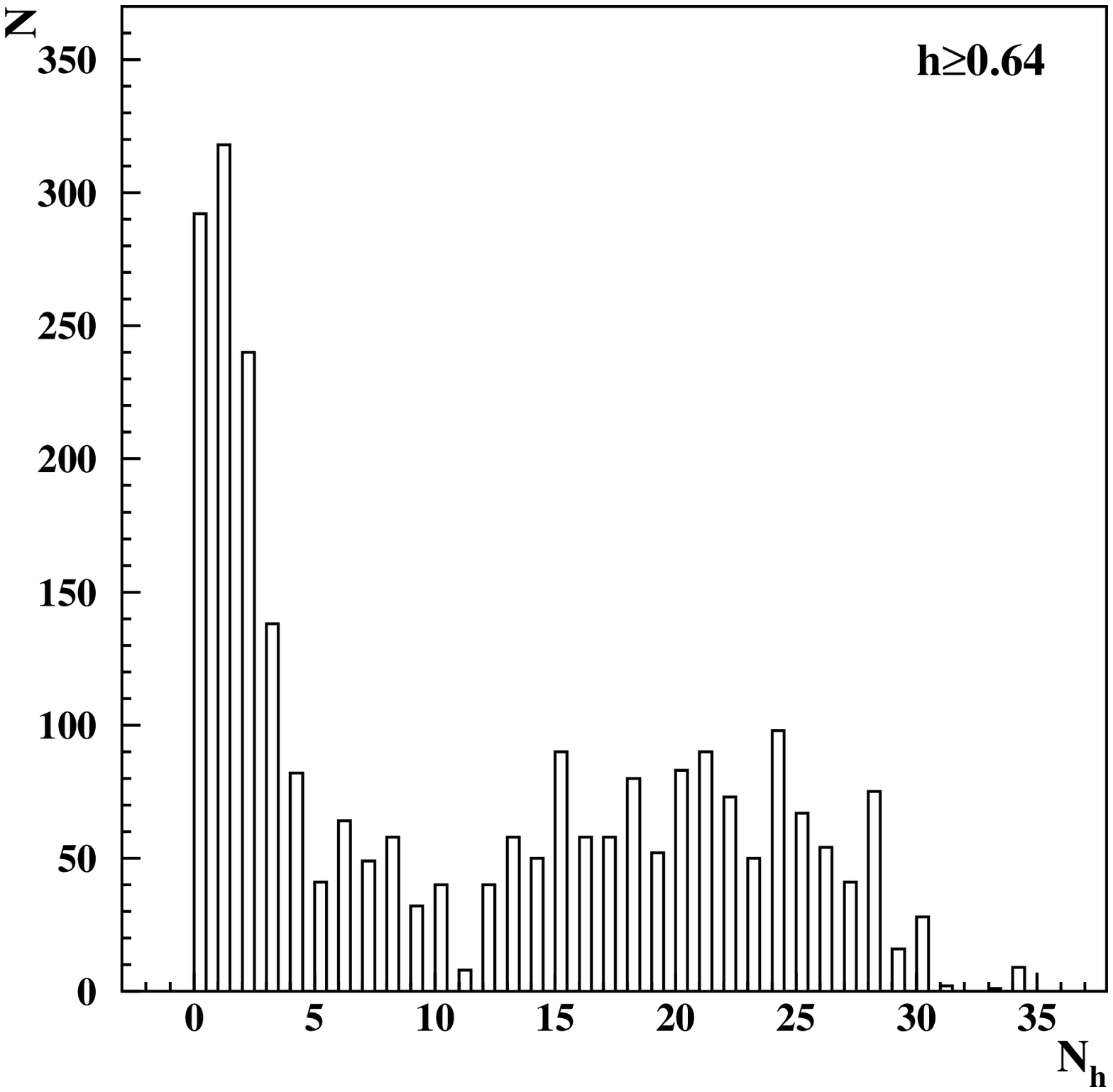}
\includegraphics*[width=0.48\textwidth,angle=0,clip]{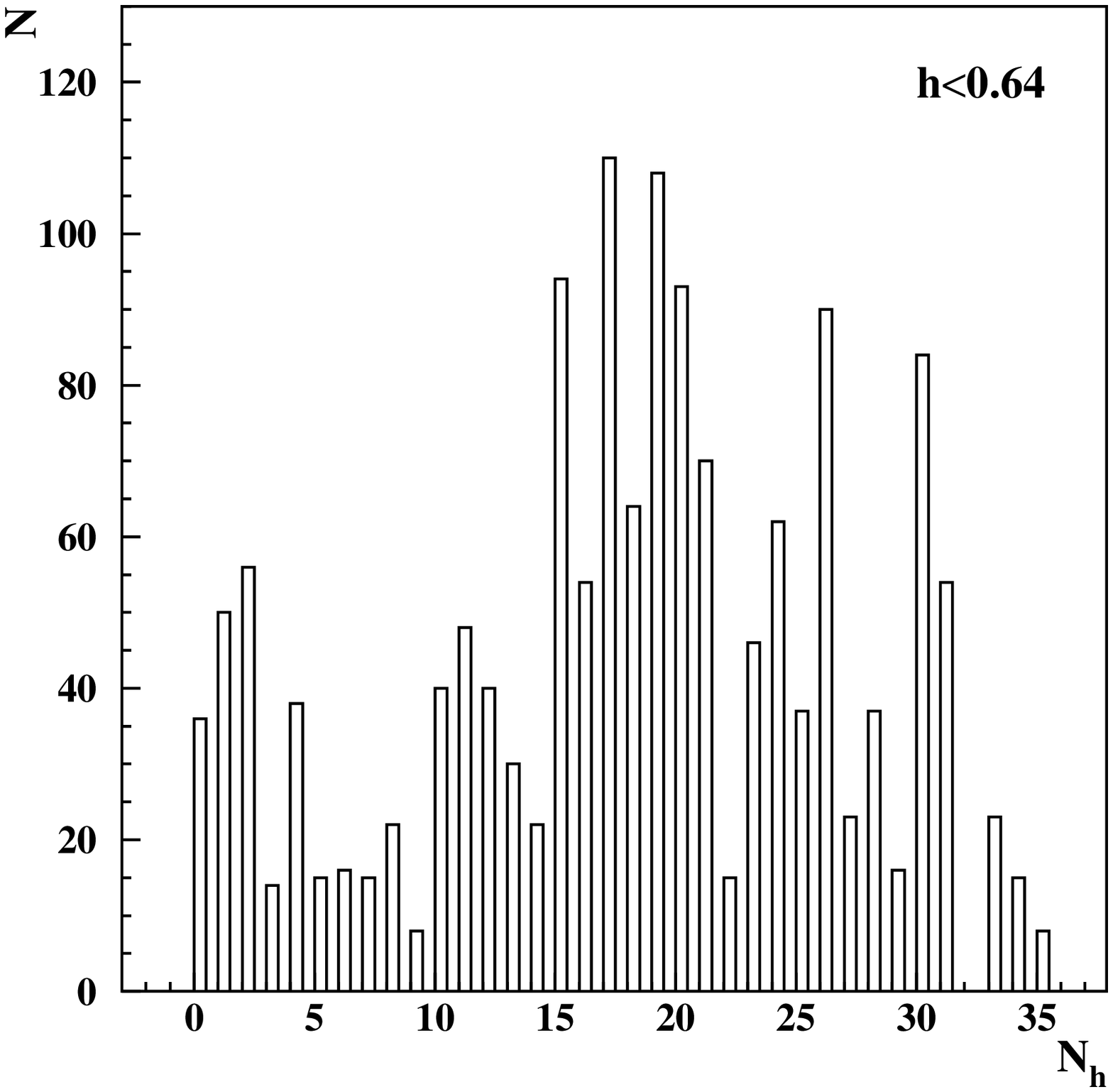}
\caption{\label{fig1} 
$N_h$-distribution of  target-nuclei fragments for events of Au+Em 10.7 AGeV with Hurst index $h\ge0.64$ (left) and with $h <0.64$ (right).}
\end{center}
\end{figure}

     As it is seen from the Figure, for correlation events the peak in the field of small values $N_h$, is observed. 
For events with $h<0.64$ the distribution maximum is located in average area of change of $N_h$. 

If to consider more correlated events, i.e. to increase criterion of selection this effect is amplified. 

For demonstration of the observation in Figure 3 the $N_h$-distribution for the events, selected by criterion $h_{av}> 0.8$, is presented.

\begin{figure}[tbh]
\begin{center}
\includegraphics*[width=0.7\textwidth,angle=0,clip]{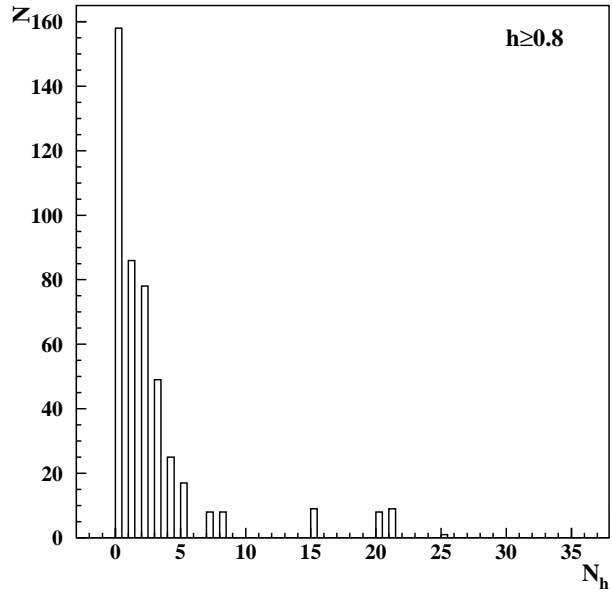}
\caption{\label{fig1} 
$N_h$-distribution of  target-nuclei fragments for events of Au+Em 10.7 AGeV with Hurst index $h>0.8$.}
\end{center}
\end{figure}

     As it is seen from the Figure, the majority of events with considerable multipartial correlations have in final distribution full destruction of  the target-nucleus.

This conclusion confirms also the general correlation distribution,  which is presented in Figure 4. Mean value of Hurst index is more for events with small number of fragments of target-nucleus.

\begin{figure}[tbh]
\begin{center}
\includegraphics*[width=0.7\textwidth,angle=0,clip]{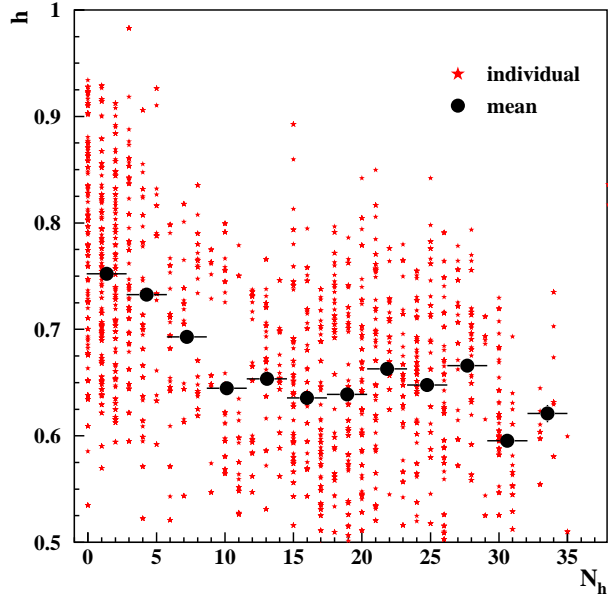}
\caption{\label{fig1} 
 the distribution of events with various values of the Hurst index versus target-nucleus fragment number $N_h$ for individual events of interaction of nuclei of  $^{197}Au$ at 10.7 AGeV with photoemulsion nuclei.}
\end{center}
\end{figure}

The following important question, which is necessary to solve: whether correlation events is connected with certain asymmetry of interactions or not. Nuclear photoemulsion represents a certain set of light, middle and heavy nuclei. It allows to analyze various types of asymmetry of the nuclear interactions, which are received in absolutely identical experimental conditions. At the same time, of course, it brings also additional problems with identification of the target-nucleus. 

In considered experiments EMU01-collaboration ($^{197}Au+Em$ 10.7 AGeV) it was used standard nuclear emulsion of type BR-2. It includes hydrogen (39.2 \%), CNO-group nuclei (35.3 \%) and nuclei of AgBr (25.5 \%). 

To separate events, which is connected with different asymmetry of interactions, we have analysed dependence of nucleus-target  
fragments number $N_h$ versus multiplicity $n_s$ for individual events of interaction of $^{197}Au$ with energy 10.7 AGeV with 
light and heavy nuclei of photoemulsion, which is presented in Figure 5. 

\begin{figure}[tbh]
\begin{center}
\includegraphics*[width=0.7\textwidth,angle=0,clip]{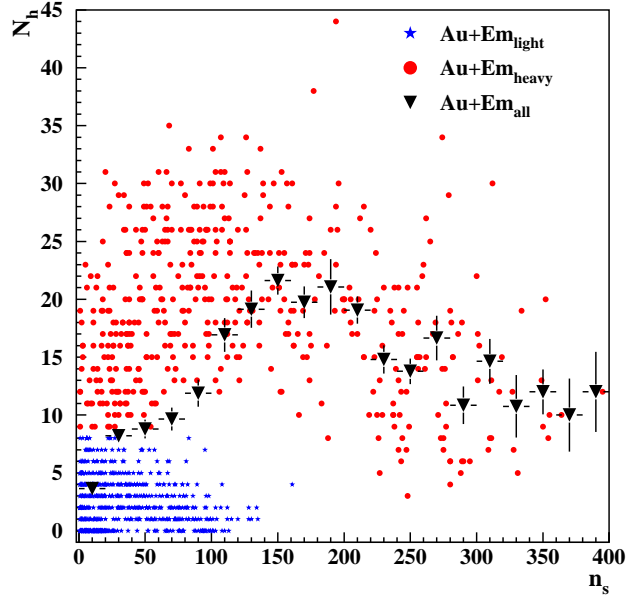}
\caption{\label{fig1} 
the distribution of events with various target-nucleus fragment number $N_h$ versus multiplicity $n_s$ for individual events of $Au+Em$ at 10.7 AGeV.}
\end{center}
\end{figure}

As it follows from Figure 5, events of interaction with light and heavy photoemulsion nuclei are separated well enough. In case of  interactions with light nuclei the choice is limited by two conditions. First, the maximum number of fragments of target-nucleus cannot exceed 8, that corresponds to largest charge of light photoemulsion nucleus - to oxygen nucleus. Secondly, the maximum multiplicity $n_s$ in interactions with light photoemulsion nuclei is much lower in comparison with interactions with heavy photoemulsion nuclei. Use of these facts allows to separate heavy events, which have large multiplicities and number of target-nucleus fragments less than 8, from light events.

On the basis of the received separation, in Figure 6 the dependence of multiplicity $n_s$ and Hurst index $h$ for individual events of interaction of $^{197}Au$ 10.7 AGeV with heavy and light photoemulsion nuclei, is presented.

\begin{figure}[tbh]
\begin{center}
\includegraphics*[width=0.7\textwidth,angle=0,clip]{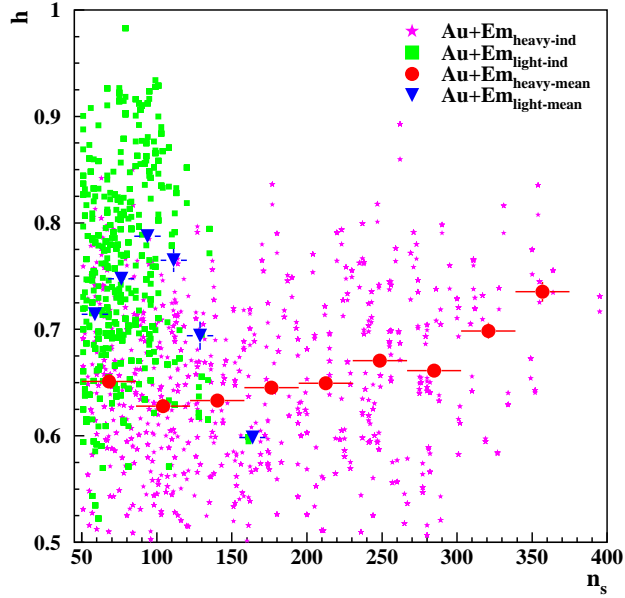}
\caption{\label{fig1} 
the distribution of events with various values of the Hurst index versus  $n_s$ multiplicity for individual events of interaction of nuclei of  $^{197}Au$ at 10.7 AGeV with heavy and light photoemulsion nuclei }
\end{center}
\end{figure}

Mean value of the Hurst index for interactions of gold nuclei with light photoemulsion nuclei represents a peak behaviour with a maximum in area of $n_s=100$, where events with the highest values of the $h$, are observed. 

     Hence, the most considerable multiparticle pseudorapidity correlations are discovered in the central interactions of heavy nuclei of gold and CNO-group nuclei, i.e. nuclei with strongly asymmetric  volume, nuclear mass, charge, etc.. 

    For research of characteristics of the discovered correlation events we have passed from the analysis of mean Hurst index to analysis of short-range and long-range multiparticle correlations on the basis of research of Hurst curve behaviour in each individual event. 

On the basis of the detailed analysis of individual interactions all events are divided on four types. Characteristic events of various types are presented in Figure 7.   

\begin{figure}[tbh]
\begin{center}
\includegraphics*[width=0.7\textwidth,angle=0,clip]{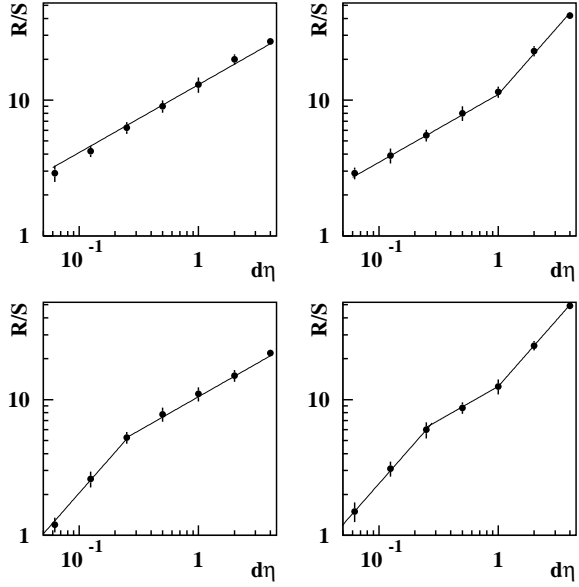}
\caption{\label{fig1} 
Behaviour of Hurst curve for four typical events of Au+Em 10.7 AGeV }
\end{center}
\end{figure}

  The first type is characterized by linear behavior of Hurst curve with $h\sim 0.5$ (Fig.7 left-upper). 
It corresponds to uncorrelated pseudorapidity distribution of secondary particles. 
Events of such type are well described on the basis of simulating with stochastic pseudorapidity distribution.  

    Events of the second type are characterised by Hurst index $h>0.64$ in the field of small pseudorapidity intervals and $h\sim 0.5$ 
at other $d\eta$(Fig.7 left-lower). 
Such behaviour of a correlation curve can be initiated short-range multiparticle correlations. It corresponds to processes of jet type. 

    Events with change of Hurst index to $h>0.64$ in the field of large values of pseudorapidity intervals have been related to 
the third type (Fig.7 right-upper). Such behaviour of a correlation curve corresponds to essential display of long-range multiparticle 
correlations and such  events are referred to processes of explosive type.  

    The events with $h>0.64$ both at small values and at large value of pseudorapidity interval, 
but with $h\sim 0.5$ in middle region of $d\eta$, have been referred to the fourth type (Fig 7. right-lower). 
Such behaviour of a curve of Hurst corresponds to events of the mixed type, including processes of explosive type and jet type.
     
Detailed analysis shown, that events of different types had different parameters of fragmentation. 

Events of the explosive type are processes of full destruction of projectile nucleus, in which multi-charge fragments are absent. 
Events with several multi-charge fragments of projectile nucleus give Hurst curves corresponding processes of mixed type.
In uncorrelated events and in processes of jet type only one multi-charge fragment is discovered. 

Also interesting results follows from joint analysis of fragmentation parameters and multiplicity distributions, 
which are presented in Figure 8. 

\begin{figure}[tbh]
\begin{center}
\includegraphics*[width=0.7\textwidth,angle=0,clip]{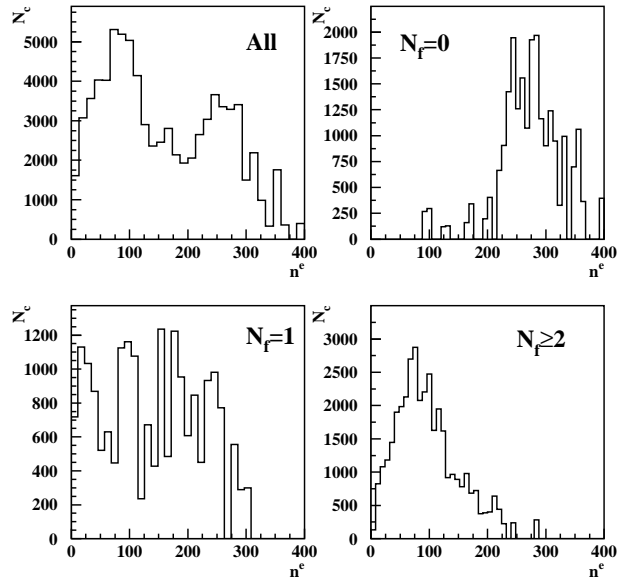}
\caption{\label{fig1} 
Multiplicative $n^e$-distribution for events with different number of multi-charge fragments in Au+Em 10.7 AGeV }
\end{center}
\end{figure}

In order to reveal a deposit of low-probability (but very important) events with large multiplicity, 
in the figure a multiplicative distribution is presented. In this approach a weight of each event is equal to particles number 
in the event, i.e. $N_c=\sum n^e$, where $n^e$ - multiplicity of individual event.       
 
As it is seen from the Fig.8, multiplicity distribution is distribution with two clear humps (Fig.8 left-upper). 

Events of explosive type, which are characterised by full absence of multi-charge fragments $N_f = 0$ (Fig. 8 right-upper), 
have large multiplicity. They form the hump with average multiplicity $<n^e>\sim 272$.

    In events of the mixed type, which have several multi-charge fragments, 
give main deposit to the hump with $<n^e>\sim 97$ (Fig. 8 right-lower). 

Events with $N_f = 1$, which are corresponding to uncorrelated events and processes of jet type, 
have different multiplicity in wide intervals without clearly expressed peculiarities. 

\section {Conclusion}

Investigation of density structure of pseudorapidity fluctuations in interactions of Au 10.7 AGeV nuclei with 
photoemulsion nuclei by Hurst method, is carried out. 
As result, events with long-range multiparticle correlations, short-range multiparticle correlations and 
mixed type in pseudorapidity distribution of secondary particles, are discovered. 

Detailed analysis shown, that events of different types had different parameters of fragmentation and specific multiplicity distributions.

      On the basis of research of dependence of multiplicity versus target-nuclei fragments number for events of various types 
it is revealed, that the most considerable multiparticle correlations are observed in interactions of mixed type, which 
correspond to the central collisions of gold nuclei and nuclei of CNO-group, i.e.  nuclei with strongly asymmetric volume, 
nuclear mass, charge, etc. Such events are characterised by full destruction of target-nucleus and disintegration of 
projectile-nucleus on several multi-charged  fragments.

\vspace{12pt}

{\Large \bf Acknowledgements }

This work was supported by grant N1563/GF of Ministry of Education and Science of Kazakhstan Republic.  


\end{document}